\pdfoutput=1 

\documentclass[
    aps,              
    prl,              
    reprint,          
    superscriptaddress, 
    longbibliography, 
    nofootinbib,      
    floatfix          
]{revtex4-2}

\usepackage[utf8]{inputenc}
\usepackage[T1]{fontenc}
\usepackage{graphicx}
\usepackage{dcolumn}
\usepackage{bm}
\usepackage{amsmath, amssymb}
\usepackage{enumitem}
\usepackage{hyperref}
\hypersetup{
    colorlinks=true,
    linkcolor=blue,
    filecolor=magenta,      
    urlcolor=blue,
    citecolor=blue,
}

\begin{document}
\title{{Real-time Scattering in \texorpdfstring{$\mathbf{\phi^4}$}{phi4} Theory using Matrix Product States}}
\author{Bahaa Al Sayegh}
    \email{alsayeghbahaa@gmail.com}
    \affiliation{Department of Physics, Lebanese University, Hadat, Lebanon}

\author{Wissam Chemissany}
    \email{wchem@sas.upenn.edu}
    \affiliation{David Rittenhouse Laboratory, University of Pennsylvania, Philadelphia, PA 19104, USA}\affiliation{Department of Physics, Freie Universit\"at Berlin, 14195 Berlin, Germany}

\date{\today}

\begin{abstract}
We investigate the critical behavior and real-time scattering dynamics of the interacting $\phi^4$ quantum field theory in (1+1)-dimensions using uniform matrix product states (uMPS) and the time-dependent variational principle (TDVP). A finite-entanglement scaling analysis at $\lambda = 0.8$ bounds the critical mass-squared to $\mu_c^2 \in ]-0.2595,-0.2594[$ and provides a quantitative map of the symmetric, near-critical, and spontaneously broken regimes. Using these ground states as asymptotic vacua, we simulate two-particle collisions in a sandwich geometry and extract the elastic scattering probability $P_{11\to 11}(E)$ and Wigner time delay $\Delta t(E)$ using a sandwich geometry protocol. We find strongly inelastic scattering in the symmetric phase ($P_{11\to 11} \simeq 0.712$, $\Delta t \simeq -158$ for $\mu^2 = +0.2$) and almost perfectly elastic collisions in the spontaneously broken phase ($P_{11\to 11} \simeq 1$, $\Delta t \simeq -108$ for $\mu^2=-0.1$ and $P_{11\to 11} \simeq 1$, $\Delta t \simeq -177.781$ for $\mu^2=-0.5$). Crucially, the scattering protocol exhibits a distinctive divergence near the critical coupling; we show that this behavior serves as a dynamical signature of the quantum critical point, arising directly from the closing of the mass gap. These results demonstrate that TDVP-based uMPS can effectively probe nonperturbative scattering and critical dynamics in lattice field theories with controlled entanglement truncation.
\end{abstract}

\maketitle

\section{Introduction}

Quantum field theory (QFT) provides the standard framework for describing relativistic many-body systems and phase transitions~\cite{peskin1995qft,lancaster2014qft,
coleman2018lectures,tong2006qft}. A paradigmatic example is the quartic interaction model, or $\phi^4$ theory~\cite{campbell2019phi4}, whose action in $(1+1)$ dimensions reads 
\begin{align}
    \mathcal{S}_{\phi^4}
    &=
    \int d^2 x\, \mathcal{L}_{\phi^4} \nonumber \\ 
    &=
    \int d^2 x\,
    \bigg[
        \frac{1}{2} (\partial_0 \phi)(\partial^0 \phi) - \frac{1}{2}(\partial_x\phi)
        (\partial^x\phi) \nonumber \\
        &~~~~~~~~~~~~~~~~~~~~~~~-
        \bigg(
            \frac{1}{2}m^2 \phi^2
            +
            \frac{g}{4!}\phi^4
        \bigg)
    \bigg],
    \label{eq:phi4_action}
\end{align}
where $\mathcal{L}_{\phi^4}$ is the Lagrangian density, $\phi \equiv \phi(x)$ is a real scalar field, $m$ is the mass parameter, and $g>0$ controls the strength of the quartic self-interaction. The $\phi^4$ model underlies the field-theoretic description of classical and quantum critical phenomena~\cite{huang1987statistical,Cardy1996Scaling,
sachdev2011qpt,kleinert2000critical} and provides a canonical realization of Ising universality in $(1+1)$ dimensions~\cite{simon1973phi4ising,
campbell2019phi4}.

Despite the success of perturbative techniques such as Feynman diagrams~\cite{feyn54,Dyson1949RadiationTheories} and renormalization-group methods~\cite{Wilson1971-RG-I}, they offer limited access to real-time dynamics and strongly coupled regimes, leaving nonperturbative information about scattering processes typically extracted from collider experiments (e.g., the SPS, Tevatron, and LHC) rather than from first-principles calculations.

In two dimensions, complementary nonperturbative insights into scattering and resonance structure can be obtained from the analytic $S$-matrix program and integrable deformations of Ising field theory~\cite{karowski1978twoparticle, McCoy:1978ta,delfino1995spinspin,Fonseca2001-IFTAnalyticFreeEnergy, Delfino2006-DecayAboveThreshold, gabai2022isingSMatrix,mizera2023analyticSmatrix}, but these approaches do not directly address real-time wave-packet dynamics in a given microscopic Hamiltonian.

In parallel, there is a growing effort to compute real-time scattering amplitudes directly on quantum computers. Algorithmic proposals for simulating scattering in scalar quantum field theories on fault-tolerant devices have been put forward in Ref.~\cite{jordan2012quantumScattering}, and recent proof-of-principle digital quantum simulations have begun to implement scattering dynamics for scalar and Ising-like field theories using up to $\mathcal{O}(10^2)$ qubits, W-state encodings, and Hamiltonian truncation on noisy intermediate-scale hardware~\cite{zemlevskiy2025-scalableScalarScattering,farrell2025-WstatesScattering,ingoldby2025-HTscattering}. While Hamiltonian truncation has successfully established high-precision benchmarks for the static phase diagram~\cite{RychkovSlavaVitaleLorenzoHAMTRUN}, tensor network methods provide a more direct route to simulating real-time dynamics in the thermodynamic limit.

In $(1+1)$ dimensions, matrix product states (MPS)~\cite{milstead2013mps,
verstraete2008mps,Orus2014-PracticalTN}, combined with the time-dependent variational principle (TDVP) and tangent-space methods~\cite{haegeman2011tdvp,haegeman2016unifying,vanderstraeten2019tangent,milsted2013variational}, provide a complementary route to real-time evolution in the thermodynamic limit. This approach has enabled simulations of false-vacuum collisions, Schwinger pair production, and gauge dynamics~\cite{Milsted2022-FalseVacuum,buyens2017,Pichler2016}, and was recently used by Jha \emph{et al.} to study real-time scattering in Ising field theory~\cite{Jha2025-IFT-MPS}.

Here we extend TDVP-based uniform MPS (uMPS) methods to the interacting $\phi^4$ field theory with the goal of probing both its critical behavior and real-time two-particle scattering across the phase diagram. We focus on a fixed self-interaction parameter and use finite-entanglement scaling (FES)~\cite{milstead2013mps} to locate the quantum critical point and characterize the adjacent phases. The FES analysis yields an estimate for the critical mass and provides a quantitative map of the symmetric, near-critical, and spontaneously broken regimes that we subsequently explore via scattering simulations.

For our computations, we require a discretized Hamiltonian. Therefore, starting with the Legendre transformation of the Lagrangian density $\mathcal{L}_{\phi^4}$ introduced in Eq.(\ref{eq:phi4_action}), we discretize the spatial dimension onto a one-dimensional lattice with spacing $a$. By approximating the spatial derivative using a finite difference, we obtain the following Hamiltonian:
\begin{equation}
    H_{\phi^4} = \frac{1}{a}\sum_j \left(
        \frac{1}{2}\pi_j^2 
        + \frac{1}{2}\left(\phi_{j+1} - \phi_j\right)^2
        + \frac{1}{2}\mu^2\phi_j^2 + \frac{\lambda}{4!}\phi_j^4        
                \right),
\label{eq:ham_phi4_discrete}
\end{equation}
with $\pi_j$ representing the conjugate momenta, $\mu^2 \equiv a^2m^2$ the bare mass-squared parameter, and $\lambda \equiv a^2g$ the quartic coupling. We can neglect the overall factor of $\frac{1}{a}$ and in our work we have set $a = c = \hbar = 1$. The system respects the canonical commutation relation $[\phi_j,\pi_k] = i\delta_{jk}$ and treats ($\mu,\lambda$) as bare lattice couplings. Local Hilbert space dimensions are truncated to a local value $d$.

We renormalize the Hamiltonian by subtracting the local ground-state energy density to ensure finite total energy. As noted by Sugihara~\cite{Takanori_Sugihara_2004}, standard normal ordering is defined in the Fock basis and is non-local in a real-space lattice representation. We therefore employ a local subtraction scheme consistent with the vacuum energy removal used in Hamiltonian truncation estimates.

To describe the state of the system we choose the MPS representation. In this framework, the lattice chain is represented by sites that carry information about the system via square $D \times D$ matrices, $A^{s_n}$, where $D$ is the maximum bond dimension of the MPS. This parameter controls the accuracy of the ansatz and sets an upper bound on the entanglement entropy, $S \le \log(D)$. 

MPS can generically take the form
\begin{equation}
    |\Psi\rangle = \sum_{\{s\}} \left( \vec{v}_L^{\dagger} \dots 
                        A_{-1}^{(s-1)} A_0^{(s_0)} A_1^{(s_1)}\dots \vec{v}_R\right) |s\rangle,
    \label{eq::MPS}
\end{equation}
where $|s\rangle = |\dots s_{-1}s_0s_1\dots\rangle$, and $\vec{v}_L$ and $\vec{v}_R$ are vectors that live at spatial infinity and are irrelevant in the thermodynamic limit.

One can see that Eq.(\ref{eq::MPS}) is invariant under gauge transformations of the form $A_n^{(s_n)}\rightarrow g A_n^{(s_n)} g^{-1}$; in particular, there exists a gauge transformation that allows all matrices to become site-independent, i.e. $A^{s_n}_n = A^{s_n}$ for all sites $s_n$. In this case, the representation is known as a uniform matrix product state (uMPS) and takes the form
\begin{equation}
    |\Psi\rangle = \sum_{\{s_i\}} \vec{v}_L^{\dagger} 
                \left( \prod_{n \in \mathbb{Z}} A^{(s_n)} \right) \vec{v}_R \; |s\rangle
\end{equation}

The central object in MPS is the transfer matrix,
\begin{equation}
    E = \sum_{s=1}^d A^s \otimes \bar{A}^s,
\end{equation}
a completely positive map whose determinant eigenvalue can be scaled to 1. In the generic case the leading eigenvalue is non-degenerate with positive left and right fixed points $l$ and $r$, normalized by $\mathrm{Tr}(lr) = 1$. These properties ensure well-defined thermodynamic quantities and independence from boundary vectors.

Local observables in this representation are evaluated by contracting transfer matrix fixed points. For some operator $O$, with $\langle O \rangle = \langle l|E_O|r\rangle$,
\begin{equation}
    E_O = \sum_{s,t}\langle s|O|t\rangle A^s\otimes\bar{A}^t.
\end{equation}
For a two-site operator, say the nearest-neighbor Hamiltonian $\langle h\rangle=\langle l|E_h|r\rangle$,
\begin{equation}
    E_h = \sum_{s_1,s_2,t_1,t_2}\langle s_1s_2|h|t_1t_2\rangle \left(A^{s_1}A^{s_2}\right)\otimes\left(\bar{A}^{t_1}\bar{A}^{t_2}\right).
\end{equation}

For one-dimensional (1D) systems, the matrix product state (MPS) ansatz serves as a high-fidelity variational manifold $\mathcal{M}_{\text{MPS}}$ that efficiently captures the required area law entanglement structure of ground and low-energy excited states. The time-dependent variational principle (TDVP) \cite{vanderstraeten2019tangent,milsted2013variational,haegeman2011tdvp} provides a rigorously grounded framework for performing both real- and imaginary-time evolution while strictly confining the state's trajectory to this low-dimensional manifold, $|{\Psi(A(t))\rangle}\in\mathcal{M}_{\text{MPS}}$.

\begin{figure}[htbp]
    \centering
    \includegraphics[width=0.8\linewidth]{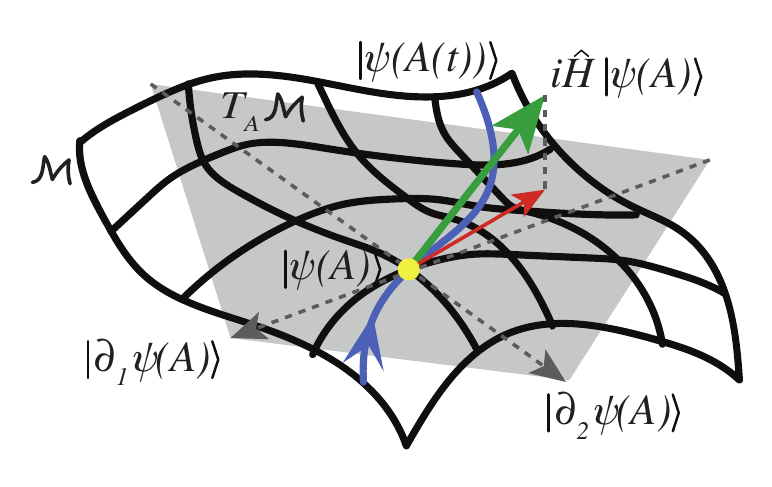}
    \caption{Time-Dependent Variational Principle as depicted in Haegeman et al.~\cite{haegeman2011tdvp}}
    \label{fig::TDVP}
\end{figure}

The TDVP procedure commences with the time-dependent Schr\"odinger equation,
\begin{equation}
    i\frac{\partial}{\partial t}|{\Psi(t)}\rangle = \hat{H}|{\Psi(t)}\rangle.
\end{equation}
While the exact evolution vector $\hat{H}|{\Psi(A)}\rangle$ generally points out of the manifold $\mathcal{M}_{\text{MPS}}$, the TDVP seeks an optimal path $|{\Psi(A(t))}\rangle$ by projecting the instantaneous evolution onto the tangent space $T_A\mathcal{M}_{\text{MPS}}$ at the current state, $|{\Psi(A)}\rangle$. This optimization is formulated as a minimization problem for the residual error vector:
\begin{equation}
    \dot{A} = 
    \mathop{\arg \min}_B \left|\left|i\hat{H}|\Psi(A)\rangle-|{\Phi(B;A)}\rangle\right|\right|^2,
\end{equation}
where $|{\Phi(B;A)}\rangle = B^i\frac{\partial}{\partial A^i}|{\Psi(A)}\rangle =B^i|{\partial_i\Psi(A)}\rangle$ is a generic tangent vector parametrized by the tensor $B$, and $A$ denotes the collective MPS variational parameters.

The minimization condition is mathematically equivalent to demanding that the residual error vector be orthogonal to the tangent space, which yields the TDVP equations of motion:
\begin{equation}
    \langle{\partial_{\bar{j}}\Psi|\partial_i\Psi}\dot{A}^i\rangle = -i\langle{\partial_{\bar{j}}\Psi|\hat{H}|\Psi}\rangle.
\end{equation}
The TDVP evolution is thus an orthogonal projection of the exact dynamics onto the MPS manifold (see Fig.~\ref{fig::TDVP}), represented by the flow equation:
\begin{equation}
    i\frac{\partial}{\partial t}|{\Psi(A(t))}\rangle = P_{A(t)}\hat{H}|{\Psi(A(t))}\rangle,
\end{equation}
where $P_A$ is the orthogonal projector onto $T_A\mathcal{M}_{\text{MPS}}$.

In this work we first leverage the scaling of entropy with bond dimension, noting that a finite $D$ imposes an effective correlation length $\xi_D$ even at criticality. Near a conformal fixed point, the half-chain von Neumann entropy $S$ obeys the finite-entanglement scaling relation
\begin{equation}
    S \simeq \frac{c}{6} \log{\xi_D} + \text{const.},
    \label{eq:FES}
\end{equation}
with $c$ the central charge and $\xi_D \propto D^\kappa$. This relation enables us to estimate the critical coupling $\mu^2_c$ and verify the Ising universality class directly from uMPS ground states, without changing system size. The resulting bracket for $\mu_c^2$ then serves as input for our real-time scattering simulations in the different phases.

In a second step, we use these uMPS ground states as asymptotic vacua for real-time scattering simulations in a nonuniform ``sandwich'' geometry. Following the prescription of Jha \textit{et al.}~\cite{Jha2025-IFT-MPS}, we imprint two localized wave packets with opposite momenta on top of the uniform background and evolve the resulting state using TDVP for a finite window embedded in the bulk. From the asymptotic outgoing state we extract the elastic scattering probability $P_{11\to 11}(E)$ and Wigner time delay $\Delta t(E)$, and track their behavior as the theory is tuned from the symmetric phase through the critical region into the spontaneously broken regime. This combination of finite-entanglement scaling and real-time scattering provides a unified, nonperturbative view of criticality and quasiparticle dynamics in lattice $\phi^4$ theory.

\section{Methods}

\subsection{Finite-entanglement scaling}
As discussed in the Introduction, a uniform MPS (uMPS) with finite bond dimension $D$ induces an effective correlation length $\xi_D$ and the entropy–length relation~Eq.(\ref{eq:FES}) near a conformal fixed point. Here we describe how we use this relation in practice to locate the critical coupling and test universality.

We parametrize the theory by \(r \equiv \mu^2\), and denote the critical value by $r_c \equiv \mu_c^2$.

For a fixed quartic coupling $\lambda$ we compute translationally invariant ground states $|\psi_D(r)\rangle$ for a set of bond dimensions $\{D\}$ and a sweep of mass parameters $r$ across the critical region.
For each pair $(r,D)$ we extract the correlation length $\xi_D(r)$ from the subleading eigenvalue of the uMPS transfer matrix and obtain the bipartite entropy $S_D(r)$ from the Schmidt spectrum. We then fit, at fixed $r$, $S_D(r)$ as a function of $\log \xi_D(r)$ to a linear form \(S_D(r) \simeq a(r)\,\log \xi_D(r) + b(r),\) and interpret the slope as an effective central charge, $c_{\mathrm{eff}}(r) = 6\,a(r)$.

The resulting $c_{\mathrm{eff}}(r)$ profile exhibits a convergence to $c_{\mathrm{eff}} = 0.5$ as $r$ approaches criticality. We identify a narrow interval in $r$ that contains this value and take it as a bracket for the critical coupling,
\begin{equation}
  r_c \in [r_{\min}, r_{\max}],
\end{equation}
with the peak position yielding a central estimate and variations of the fit window and $D$-range providing an uncertainty. As a complementary consistency check we also test data collapse of a local observable (e.g.\ the order parameter) when plotted as a function of $(r - r_c)\,\xi_D^{1/\nu}$, with $\nu = 1$ (Ising). This confirms that the extracted $r_c$ is compatible with Ising universality within our accessible bond dimensions. The FES analysis thus yields a quantitatively controlled estimate of the critical mass-squared and a map of the surrounding parameter space that we use to select representative points in the symmetric, near-critical, weakly broken, and deeply broken regimes.

\subsection{Real-time scattering protocol}
\begin{figure}[htbp]
    \centering
    \includegraphics[width=\columnwidth]{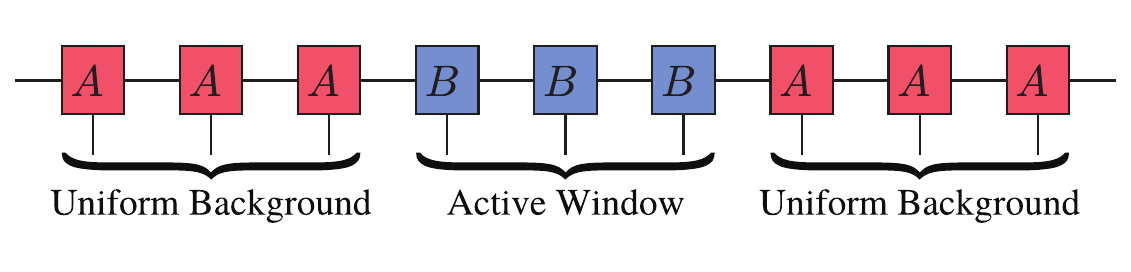}
    \caption{Sandwiching B-Tensor Inside a Uniform Ground State}
    \label{fig::SandwichB}
\end{figure}

To probe real-time scattering we follow the sandwich geometry introduced in Ref.~\cite{Jha2025-IFT-MPS}. For each choice of mass parameter $r$ at a fixed quartic coupling $\lambda$, a fixed maximal bond dimension $D$, and a fixed Hilbert space truncation $d$ we take the corresponding uMPS ground state as an approximation to the asymptotic vacuum. Starting from said uMPS ground state, we create a non-uniform `sandwich' MPS of length $N$ embedded in that uniform background, see Fig.~\ref{fig::SandwichB}. This gives a window where we can imprint and evolve localized excitations while the far left/right halves remain vacuum-like.

To build these clean quasiparticle wavepackets, we construct a tangent-space vector $B$ from the local physical operator $\phi$ by projecting $O \cdot A$ onto the left-gauge orthogonal complement of the uMPS manifold:
    \begin{equation}
        V^s = \sum_{s'}O_{ss'}A^{s'}\;,~~~B=V-A\cdot \left(\sum_sA^{s\dagger}V^s\right),
    \end{equation}
then normalize $B$. This enforces $\sum_sA^{s\dagger}B^s=0$ so the packet has no vacuum component in the chosen gauge.

We imprint a packet by adding small copies of $B$ across the window with a Gaussian envelope around a center $n_0$ and a plane-wave phase $e^{i\kappa(n-n_0)}$. This is done gently (small amplitude), in tiles, restoring canonical form and backing off if a local update fails---so the state stays well-conditioned. We do this twice to produce two-counter propagating packets with $\pm\kappa$. 
        
In short, we build $B = P_\perp [\phi \cdot A]$ and then imprint Gaussian $\pm\kappa$ packets.

We monitor the local field expectation value $\langle \phi_n(t)\rangle$ and its connected correlators throughout the collision. 

To extract scattering data, we use a script that projects the evolving state onto a compact ``two-packet'' subspace and reads off a set of diagonal Fourier amplitudes $A_{k,-k}(t)$ as mentioned in \cite{Jha2025-IFT-MPS}. Practically, it focuses on lattice sites within Gaussian windows around the instantaneous centers of the left- and right-moving packets and enforces a minimum separation $\Delta n$ so that only well-separated pairs are considered. Using a reusable reference sandwich initially in the vacuum-like state, it inserts tiny perturbations $\varepsilon B$ at selected $(i_L,i_R)$ site pairs to define basis vectors for this two-$B$ subspace. The algorithm then computes overlaps with the current state, multiplies by the plane-wave phases to target momenta $(k,-k)$, accumulates the results across the chosen pairs, and finally normalizes by $\varepsilon^2$ to stay in a linear-response regime. To keep the cost and variance under control, pairs with negligible Gaussian weight are dropped and the total pair count is capped. We mainly are concerned with time delay and probability of elastic scattering.

The time delay measures the change in arrival (or departure) time of scattered wave packets compared with non-interacting particles. Physically, it represents how long the interacting particles ``linger'' in the scattering region because of interaction effects encoded in the phase of the elastic S-matrix element $S_{11\rightarrow 11}(E)$. A positive time delay indicates that the particles spend more time interacting (as if temporarily trapped, e.g., near a resonance), while a negative delay means they pass through faster than in free motion.

We calculate here the Wigner time delay, given by:
\begin{equation}
    \Delta t=-i\partial_E \log{S_{11\rightarrow11}(E)}.
\end{equation}
To do so from overlaps at momentum $\kappa_0$, we consider nearby momenta $\kappa_\pm=\kappa_0\pm\delta\kappa$ and compute:
\begin{align}
    \phi(t) 
    &= \arg\left(\frac{\langle{\kappa_+,-\kappa_+|\psi(t)}\rangle}{\langle{\kappa_+,-\kappa_+|\psi(0)}\rangle}\frac{\langle{\kappa_-,-\kappa_-|\psi(0)}\rangle}{\langle{\kappa_-,-\kappa_-|\psi(t)}\rangle}\right) \nonumber \\ 
    &= 2(E_{\kappa_+}-E_{\kappa_-})t+\Delta(\kappa_+) + \Delta(\kappa_-),
\end{align}
this quantity captures the relative phase-shift between components of $|{\psi}\rangle$ with momenta $\kappa_{\pm}$. The first term in $\phi(t)$, due to an expected phase shift from the free theory, is subtracted, and  $\Delta(\kappa_\pm)$ is expanded to first order around $\kappa_0$:
\begin{align}
\frac{\Delta(\kappa_+)-\Delta(\kappa_-)}{2(E_{\kappa_+}-E_{\kappa_-})} &\approx\partial_E\Delta(E) \nonumber \\
&=-i\partial_E \log{S_{11\rightarrow11}(E)}.
\end{align}

As a simple estimate of the elastic scattering $1+1\rightarrow 1+1$ probability, the script compares diagonal spectral weight before and after the collision within a narrow band around $\kappa_0$. It sums $|A_{\kappa,-\kappa}(t)|^2$ over the sampled $\kappa$ values at an early time and at a late time and takes their ratio as $P$
\begin{equation}
    P_{11\rightarrow11} = \frac{|A_{\kappa,-\kappa}(t_f)|^2}{|A_{\kappa,-\kappa}(t_i)|^2}.
\end{equation}
    
We practically calculate the probability of finding particles of the `same kind' in the $t_f$ state as those in the $t_i$ state.

This ``diagonal-only'' approximation is intentionally lightweight: it avoids the expense of reconstructing the full two-momentum distribution $|A_{\kappa,\kappa'}|^2$ while still providing a stable, interpretable signal for predominantly elastic scattering. The trade-off is that genuinely off-diagonal contributions with $\kappa'\ne \kappa$ are ignored, so the estimate is most reliable when the packets are narrow in $\kappa$ and the scattering remains nearly elastic.

All ground-state optimizations and real-time evolutions are implemented using a modernized version of the open-source \textsc{evoMPS} code developed by Milsted~\cite{evoMPS}, which we have extended to support the $\phi^4$ Hamiltonian and the scattering diagnostics used here.

\section{Results}

\subsection{Validation against Known Limits}

Before investigating finite entanglement scaling or real-time scattering, we validate our uMPS framework and Hamiltonian discretization by reproducing known regimes of the $(1+1)$D $\phi^4$ theory: the free-particle dispersion, the location of the critical point, the verification of the $\beta$-parameter, and the characterization of the broken phase.

\subsubsection{Single-Particle Dispersion}
\begin{figure*}[htbp]
\includegraphics[width=0.85\linewidth]{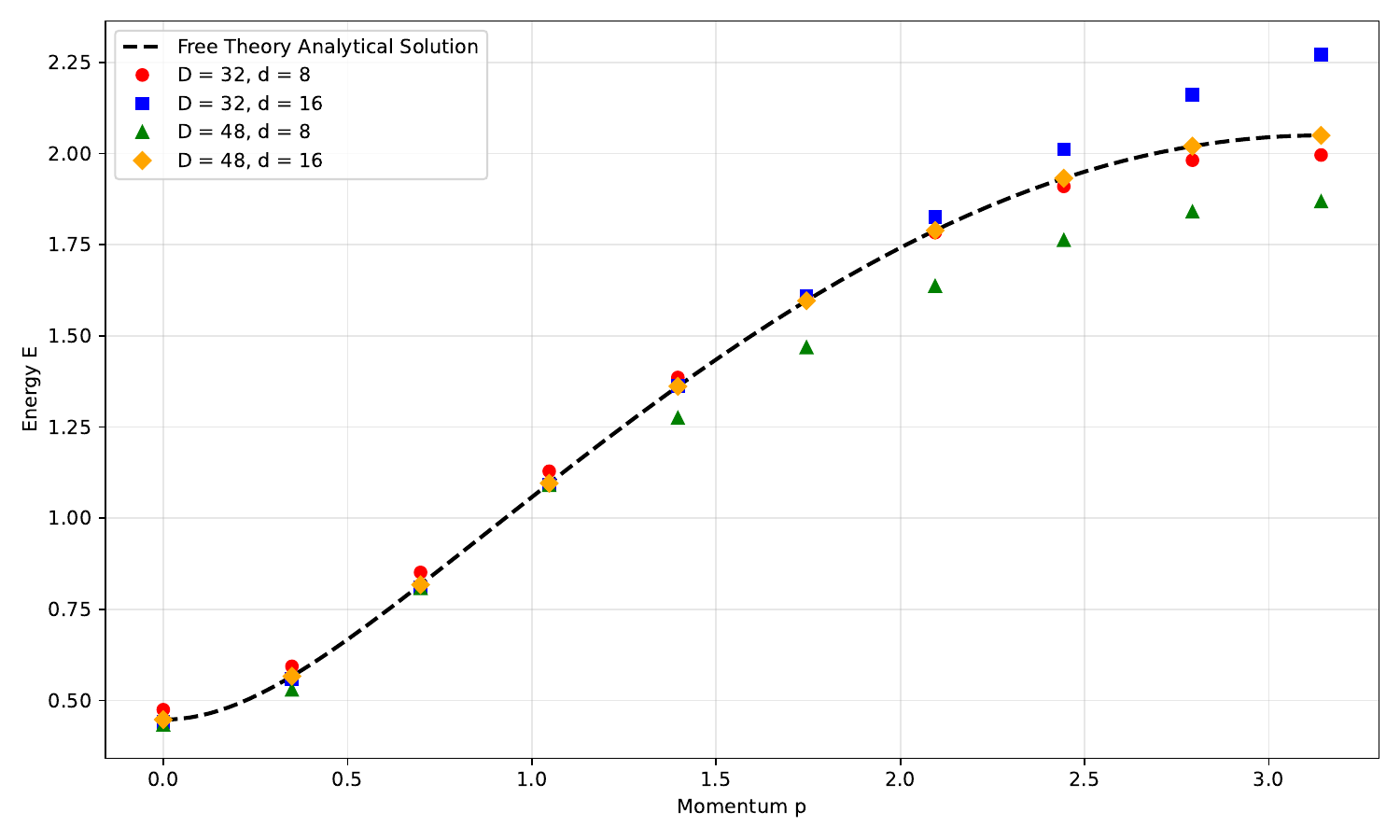}
\caption{\label{fig:dispersion} Verification of the single-particle dispersion relation in the free theory ($\lambda=0,\;\mu^2=0.2$). The MPS excitation spectra (for different bond dimensions and Hilbert-space truncations) match closely the analytic lattice dispersion (black dashed line).}
\end{figure*}

To verify that our ansatz correctly captures the kinetic terms and momentum eigenstates, we computed the excitation spectrum in the symmetric phase. Fig.~\ref{fig:dispersion} compares the MPS-derived excitation energies $E(p)$ against the analytic lattice dispersion relation for a massive boson, 
\begin{equation}
E^2(p) = \mu^2 + 4\sin^2(p/2).
\end{equation}
The simulation data shows a good agreement with the analytic curve (dashed black line) across the Brillouin zone. This confirms that our bond dimension $D=32$ and our Hilbert space truncation $d=8$ are sufficient to resolve the single-particle band without introducing significant lattice artifacts or finite-entanglement distortions.

We have taken here $\mu^2 = 0.2$.

\subsubsection{Critical Point Location}
We locate the quantum critical point (QCP) by monitoring the mass gap $\Delta$ as a function of the quartic coupling $\lambda$ at fixed $\mu^2 = -1.0$. We have taken $D = 32$ and $d = 12$.  

\begin{figure*}[htbp]
\includegraphics[width=0.85\linewidth]{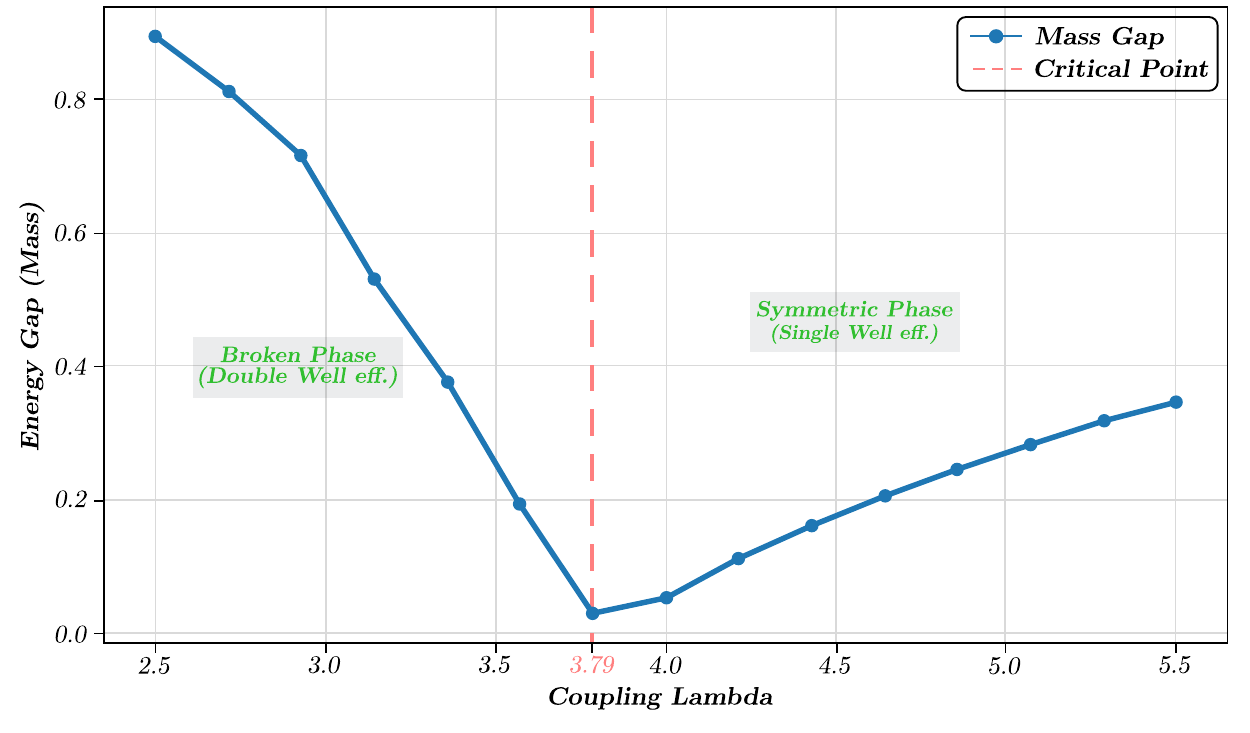}
\caption{\label{fig:criticality} Detection of the phase transition via gap closing. The energy gap (blue line) vanishes at a critical coupling $\lambda_c \approx 3.79$ (for fixed $\mu^2 = -1.0$), marking the transition from the symmetry-broken phase to the symmetric phase.}
\end{figure*}

The $\phi^4$ transition in (1+1) dimensions belongs to the 2D Ising universality class. In this class, the correlation length $\xi$ diverges at the critical point as
\begin{equation}
    \xi \propto \frac{1}{|\lambda - \lambda_c|^\nu},
\end{equation}
where $\nu = 1$ for this class. The mass (energy) gap $\Delta$ is inversely proportional to the correlation length and so
\begin{equation}
    \Delta \sim |\lambda - \lambda_c|.
\end{equation}
As shown in Fig.~\ref{fig:criticality}, the system exhibits a clear ``V-shape'' gap closing, minimizing at $\lambda_c \approx 3.79$. This closing separates the symmetric phase (single-well effective potential) from the broken phase (double-well potential), demonstrating the method's ability to capture the long-range correlations associated with the phase transition.

\subsubsection{The `Sugihara Ratio' Test}
As a further check for the accuracy of our uMPS simulations, we look at the work of T. Sugihara~\cite{Takanori_Sugihara_2004}. 
\begin{figure}[htbp]
    \centering
    \includegraphics[width=0.5\linewidth]{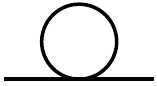}
    \caption{The divergent diagram in (1+1)D $\phi^4$ model.}
    \label{fig::tadpoleTerm}
\end{figure}
In the $\phi^4$ model, there is only one diagram that diverges in the continuum limit $a \rightarrow 0$ (see Fig.~\ref{fig::tadpoleTerm}). We can renormalize it by redefining the mass parameter as follows:
\begin{equation}
    \mu^2 = \mu^2_{\mathrm{ren}} - \delta\mu^2,
    \label{eq::mu_sq_mu_ren_delta_mu}
\end{equation}
where $\delta\mu^2$ is the tadpole counter-term that cancels the divergence and is defined as:
\begin{equation}
    \delta\mu^2 = \frac{\lambda}{2\pi\sqrt{\mu_{\mathrm{ren}}^2+4}}K\left(\frac{4}{\mu^2_{\mathrm{ren}} + 4}\right),
\end{equation}
with $K(x)$ is the complete elliptic integral of the first kind. 

Sugihara confirms, using density matrix renormalization group (DMRG) calculations, that the ratio
\begin{equation}
    R_c = \frac{\lambda}{\mu_{\mathrm{ren}}^2}\bigg|_\mathrm{critical~point} = 59.89 \pm 0.01.
\end{equation}

Table~\ref{tab::RcVerif} presents a convergence study comparing our simulated results, $R_{uMPS}$, against this theoretical target across various parameters. Because any numerical simulation using Matrix Product States is constrained by bond dimension $D$ and the local Hilbert space truncation $d$ we expect a measurable deviation from $R_c$.

For example, at a fixed coupling of $\lambda = 3.0$, increasing the bond dimension from $D=32$ to $D=64$ reduces the deviation from $4.43\%$ to $3.32\%$, demonstrating that our results systematically approach the continuum benchmark as computational power increases. The larger $11.30\%$ deviation observed at $\lambda = 3.79$ highlights the increased numerical challenge at higher interaction strength. Ultimately, these results confirm that our uMPS framework remains well within an acceptable range of theoretical expectations, providing a stable foundation for our subsequent real-time scattering analysis.
\begin{table}
    \centering
    \begin{tabular}{|c|c|c|c||c|c|c|}
        \hline
       $D$& $d$ &$\lambda$ & $\mu^2$ & $\mu^2_{\mathrm{ren}}$ & $R_{\mathrm{uMPS}}$ & Deviation from $R_c$ \\ \hline  \hline
       $32$ & $18$ & $3.79$ & $-1.0$ & $0.05686$ & $66.66$ & $11.30\,\%$ \\ \hline
       $64$ & $12$ & $3.0$ & $-0.792$ & $0.05181$ & $57.9$ & $3.32\,\%$ \\ \hline
       $32$ & $12$ & $3.0$ & $-0.809$ & $0.04801$ & $62.48$ & $4.32\,\%$ \\ \hline
       $32$ & $8$  & $0.8$ & $-0.25945$ & $0.01243$ & $64.36$ & $7.46\,\%$ \\ \hline
       
    \end{tabular}
    \caption{Comparison between the ratio $R_{\mathrm{uMPS}}$ calculated from uMPS simulations and $R_c$ from~Ref.~\cite{Takanori_Sugihara_2004}. We notice some deviation that is dependent on the bond dimension and the Hilbert space truncation, however, it lies within an acceptable range.}
    \label{tab::RcVerif}
\end{table}

\subsubsection{Symmetry Breaking and Quantum Fluctuations}
In the spontaneously broken phase, quantum fluctuations are expected to suppress the order parameter $\langle \phi \rangle$ relative to the classical mean-field prediction. We tested this by simulating the system with a small pinning field ($h=0.1$) to select a vacuum. Here we have taken $\mu^2 = -4.0$, $D = 32$, and $d = 24$.

\begin{figure*}[htbp]
\includegraphics[width=\textwidth]{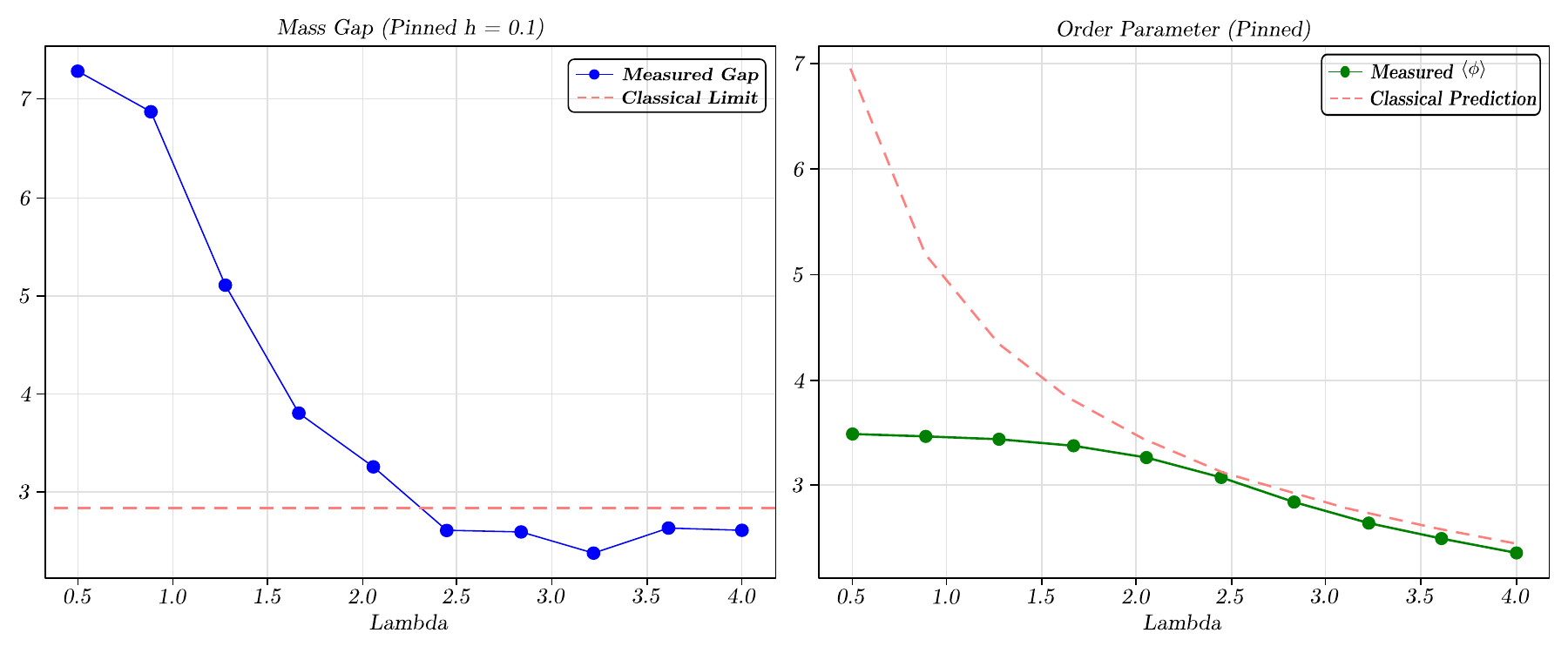}
\caption{\label{fig:broken_phase} Characterization of the broken phase with a pinning field $h=0.1$. \textbf{(Left)} The measured mass gap (blue) compared to the classical limit. \textbf{(Right)} The vacuum expectation value $\langle \phi \rangle$ (green) is suppressed relative to the classical prediction (red dashed), highlighting the method's ability to capture quantum fluctuations.}
\end{figure*}

Fig.~\ref{fig:broken_phase} (left panel) shows a convergence to the classical value of $\sqrt{2|\mu^2|}$ at high lambdas although it drifts away at low lambdas. This is because at low lambda the true vacuum moves very far out requiring a bigger Hilbert space size $d$ to capture the precise mass gap. The right panel exhibits similar behavior, it approaches the classical limit, $1/\lambda$, for higher values of lambda.

\subsection{Finite-entanglement scaling}
Using uMPS with $\lambda = 0.8$ and bond dimensions $D =\{36,48,64\}$, we performed a coarse-grained scan of the mass-squared parameter $r \equiv \mu^2$ over a broad interval and identified an approximate critical region around $r \simeq -0.259$. We then carried out a refined scan in the window $r \in [-0.261,-0.258]$ using the same FES procedure. For each pair $(r,D)$ we extracted the correlation length $\xi_D(r)$ from the subleading eigenvalue of the uMPS transfer matrix and the bipartite entropy $S_D(r)$ from the Schmidt spectrum, and fitted $S_D(r)$ as a function of $\log \xi_D(r)$ to obtain the slope $a(r)$ and effective central charge $c_{\mathrm{eff}}(r) = 6\,a(r)$, as described in the Methods.

\begin{figure*}[htbp]
    \centering
    \begin{minipage}{0.48\textwidth}
        \centering
        \includegraphics[width=\linewidth]{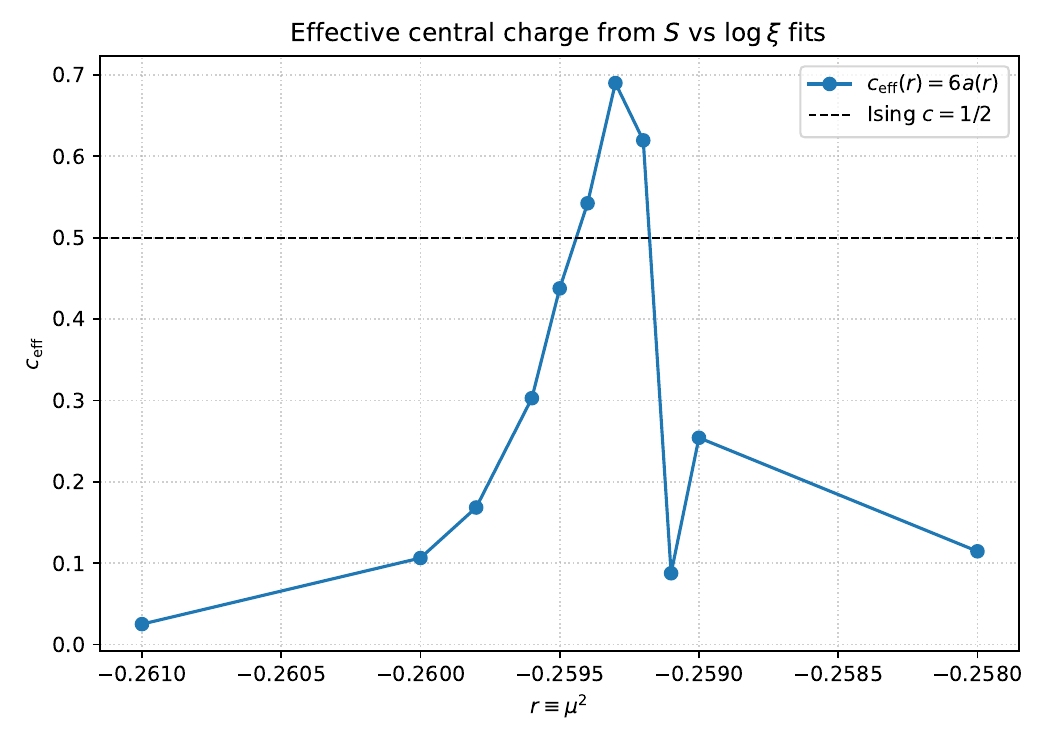}
        \vspace{0.2cm} 
        (a) $c_{\mathrm{eff}}$ vs $\mu^2$.
    \end{minipage}\hfill
    \begin{minipage}{0.48\textwidth}
        \centering
        \includegraphics[width=\linewidth]{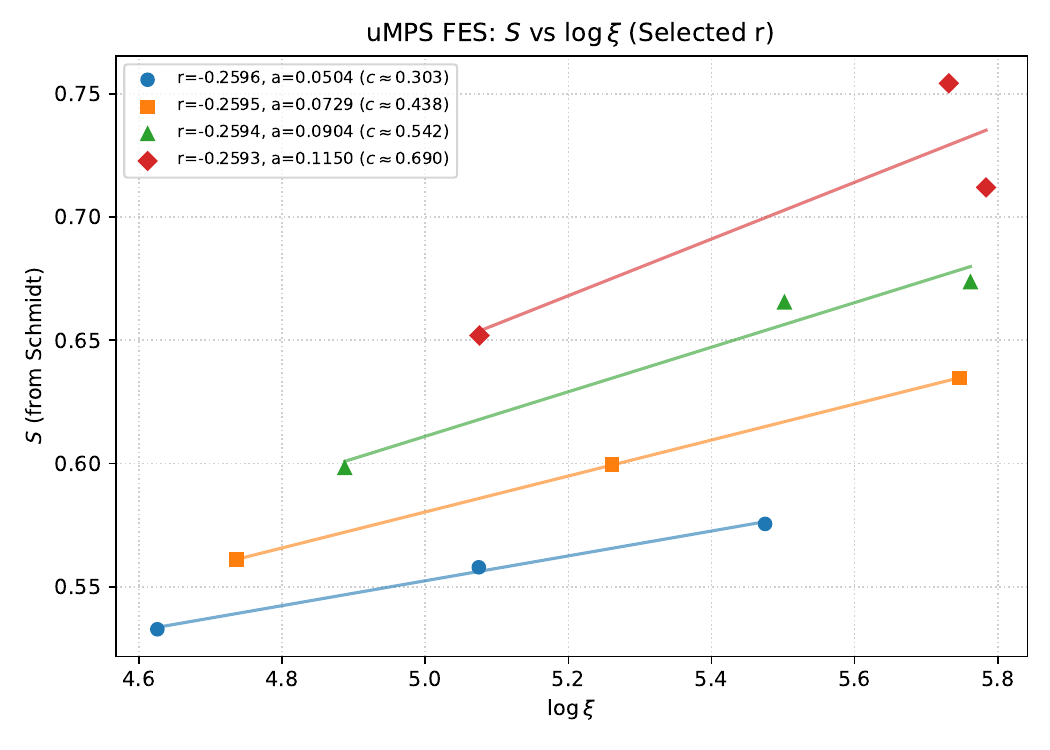}
        \vspace{0.2cm}
        (b) $\mathcal{S}$ vs $\log\xi$.
    \end{minipage}
    
    \caption{Finite-entanglement scaling near the critical point of the $\phi^4$ theory at $\lambda = 0.8$. Ground-state uMPS data with finite bond dimension $D$ are used to extract the effective central charge $c_{\mathrm{eff}}(r)$, and the entropy---correlation-length relation near $r_c = \mu_c^2$. The intersection with $c_{\mathrm{eff}} = 0.5$ and the linear dependence $\mathcal{S} \simeq \frac{c}{6} \log \xi_D + \mathrm{const.}$ close to $r_c$ yield the bracket $\mu_c^2 \in ]-0.2595,-0.2594[$ and verify Ising universality.}
    \label{fig::FESCriticality}
\end{figure*}

Across the sweep, $c_{\mathrm{eff}}(r)$ approaches $0.5$ as $r \rightarrow -0.2596$ and exhibits the following behavior in our grid [Fig.~\ref{fig::FESCriticality}(b)]:
\begin{align}
    c_{\mathrm{eff}}(-0.2596) &= 0.303 , \nonumber \\
    c_{\mathrm{eff}}(-0.2595) &= 0.438 , \nonumber \\
    c_{\mathrm{eff}}(-0.2594) &= 0.542 , \nonumber \\
    c_{\mathrm{eff}}(-0.2593) &= 0.690 , \nonumber
\end{align}
we therefore bracket the critical point as
\begin{equation}
  r_c \in ]-0.2595,-0.2594[.
\end{equation}
We will take $r_c \approx -0.25945$.

For the subsequent scattering study we use this FES map to select representative couplings. In particular, we choose a symmetric-phase point $r = +0.2$, a near-critical point close to the bracketed $r_c$, and a weakly broken point $r = -0.1$ within the spontaneously broken phase. These choices ensure that the scattering simulations sample distinct regions of the phase diagram while remaining quantitatively anchored to the FES estimate of $r_c$. A more deeply broken point at $r = -0.5$ is also included to illustrate the strongly gapped regime.

\subsection{Scattering in the \texorpdfstring{$\mathbf{\phi^4}$}{phi4} theory}
\begin{figure*}[htbp]
    \centering 
    \begin{minipage}{0.48\textwidth}
        \centering
        \includegraphics[width=\linewidth]{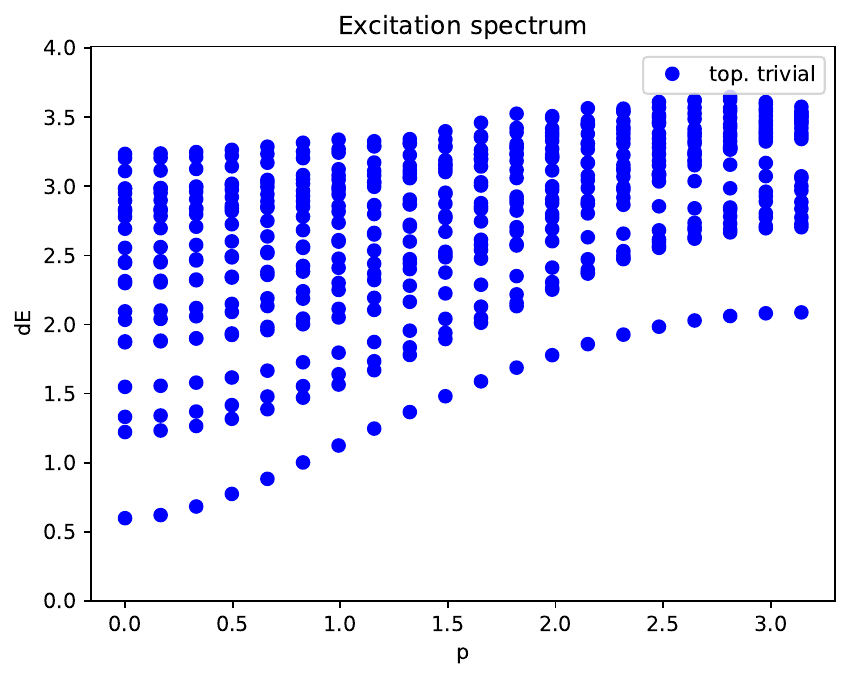}
        
        \vspace{0.2cm}
        (a) $\mu^2 = +0.2$
    \end{minipage}\hfill 
    \begin{minipage}{0.48\textwidth}
        \centering
        \includegraphics[width=\linewidth]{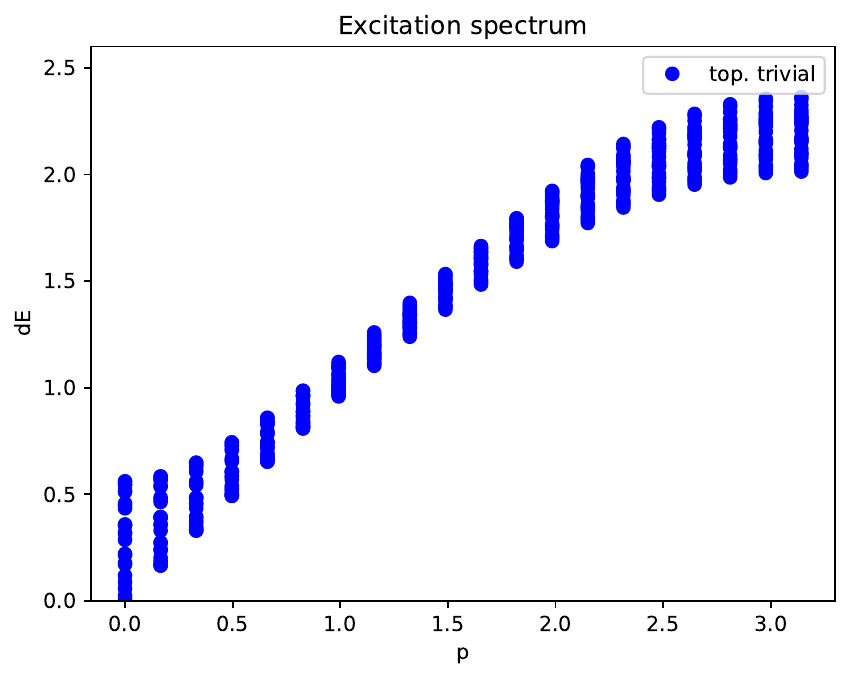}
        
        \vspace{0.2cm}
        (b) $\mu^2 = -0.2595$
    \end{minipage}
    
    \vspace{0.5cm}
    
    \begin{minipage}{0.48\textwidth}
        \centering
        \includegraphics[width=\linewidth]{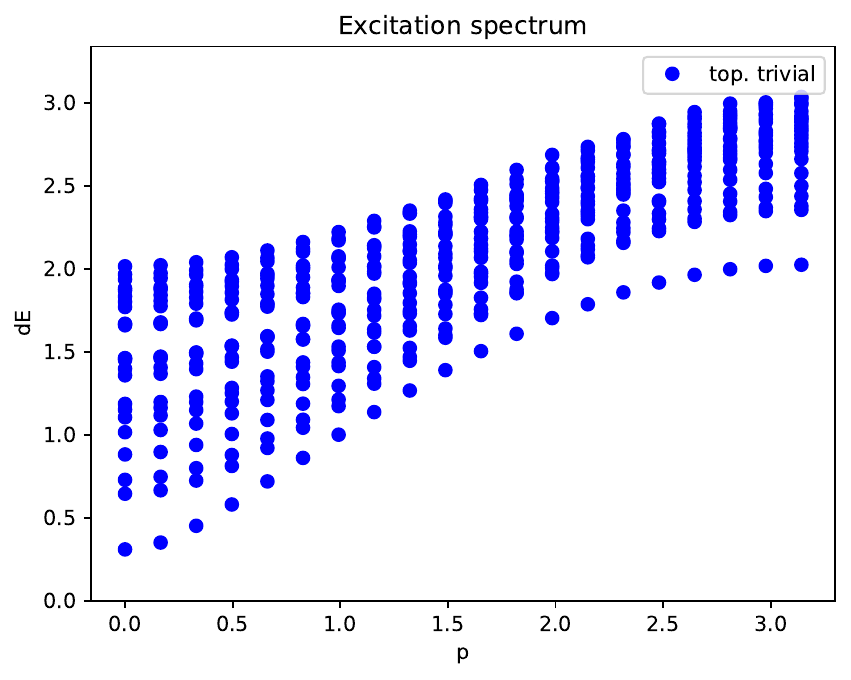}
        
        \vspace{0.2cm}
        (c) $\mu^2 = -0.1$
    \end{minipage}\hfill 
    \begin{minipage}{0.48\textwidth}
        \centering
        \includegraphics[width=\linewidth]{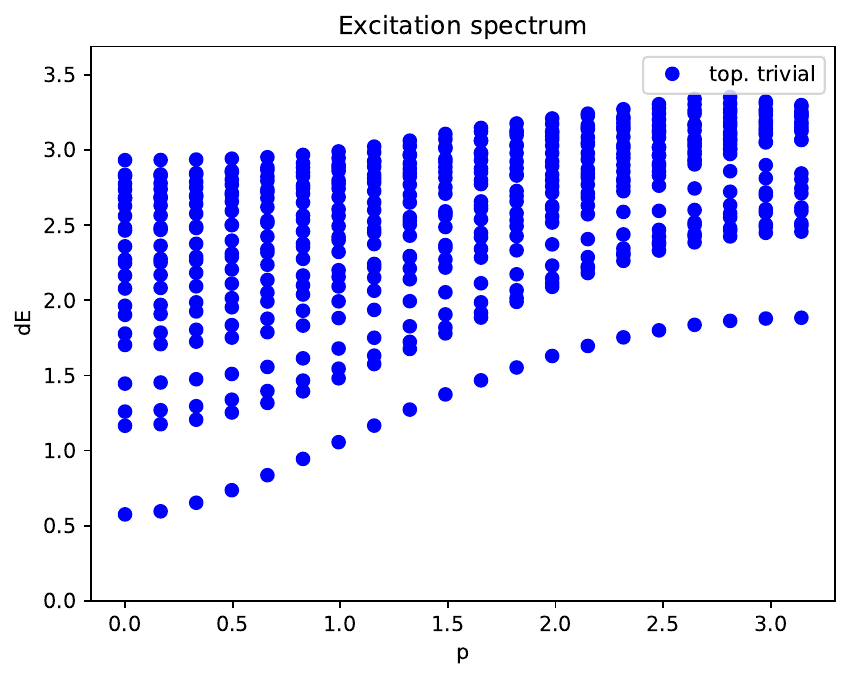}
        
        \vspace{0.2cm}
        (d) $\mu^2 = -0.5$
    \end{minipage}
    
    \caption{Excitation spectra of the $\phi^4$ theory at $\lambda = 0.8$ across the phase diagram. Each panel shows the single- and multi-particle excitation energies $d E(p)$ obtained from the uMPS transfer matrix as a function of momentum $p$ for representative values of $\mu^2$. The gap closes near (b) and reopens with a massive single-particle band in (c),(d), providing a spectral signature of the phase transition and mass generation.}
    \label{fig::spectra}
\end{figure*}

Before studying real-time dynamics, we characterize the excitation spectrum through the single-particle and multi-particle bands shown in Fig.~\ref{fig::spectra}. In the symmetric phase at $\mu^2 = +0.2$ [Fig.~\ref{fig::spectra}(a)], the system is gapped: the lowest excitation at $p=0$ has energy $d E \approx 0.6$, and there is a clear gap between the single-particle band and the multi-particle continuum, which onsets around $d E \approx 1.2$. This is characteristic of a stable symmetric phase, with a vacuum at $\phi=0$ and standard massive bosonic excitations.

Very close to the critical point at $\mu^2_c \approx -0.2594$ [Fig.~\ref{fig::spectra}(b)] the spectrum becomes gapless. The mass of the fundamental particle vanishes ($m=0$), and the lowest band starts at $d E \approx 0$, while the multi-particle thresholds (at $2m$, $3m$, etc.) collapse onto one another. The spectrum is consistent with an underlying conformal field theory: the dispersion near $p=0$ is approximately linear, $E \propto |p|$, and the stacked structure of the low-lying bands suggests a tower of stable or long-lived bound states at this coupling. 

In the spontaneously broken phase [Figs.~\ref{fig::spectra}(c), \ref{fig::spectra}(d)], the system is gapped again and a single-particle band re-emerges with a nonzero mass $m_s$. As the system moves away from criticality into the deeply broken regime (e.g., $\mu^2 = -0.5$), this mass increases, providing a spectral signature of mass generation across the phase diagram.

\begin{figure*}[htbp]
    \centering 
    \begin{minipage}{0.48\linewidth}
        \centering
        \includegraphics[width=\linewidth]{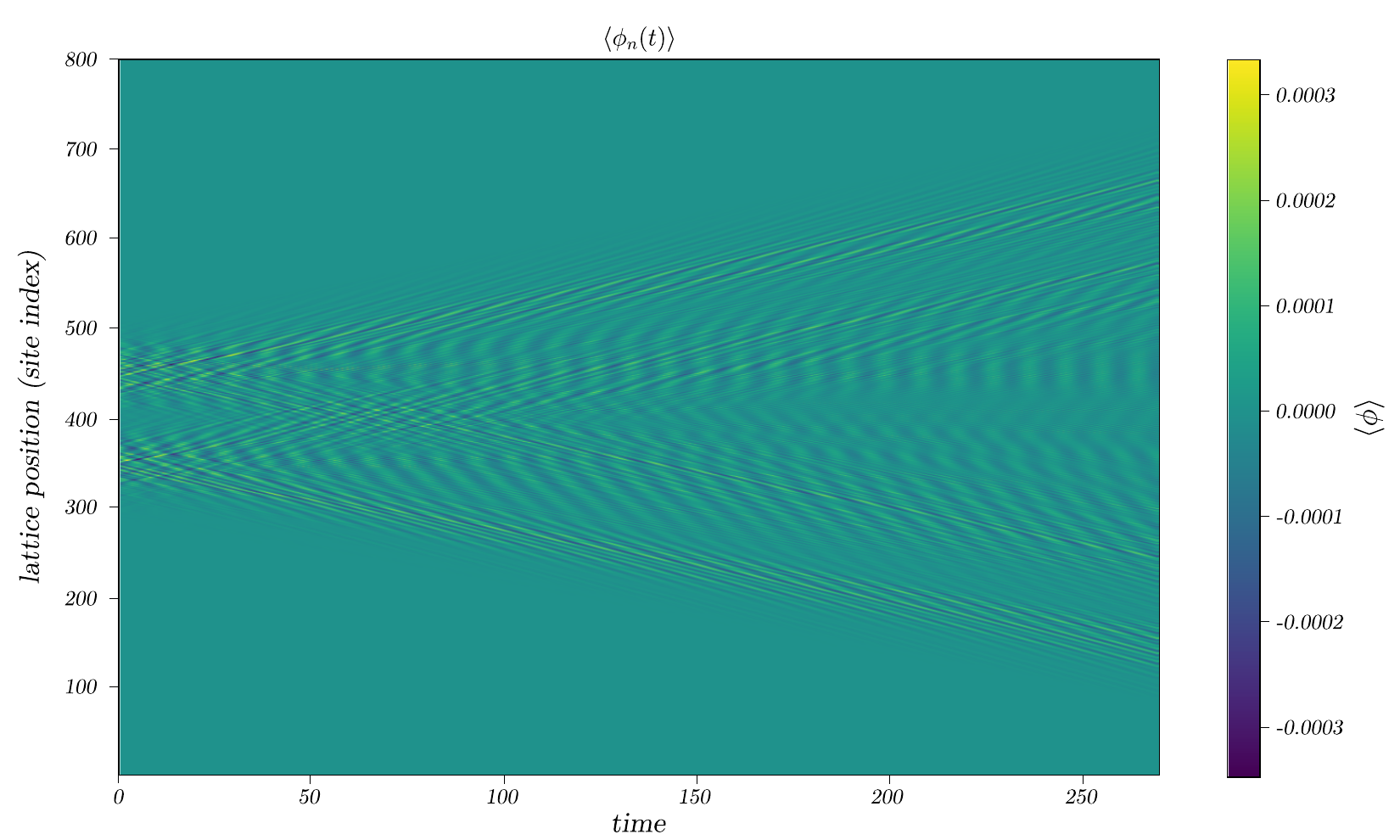}
        
        \vspace{0.2cm}
        (a) $\mu^2 = +0.2$
    \end{minipage}\hfill 
    \begin{minipage}{0.48\linewidth}
        \centering
        \includegraphics[width=\linewidth]{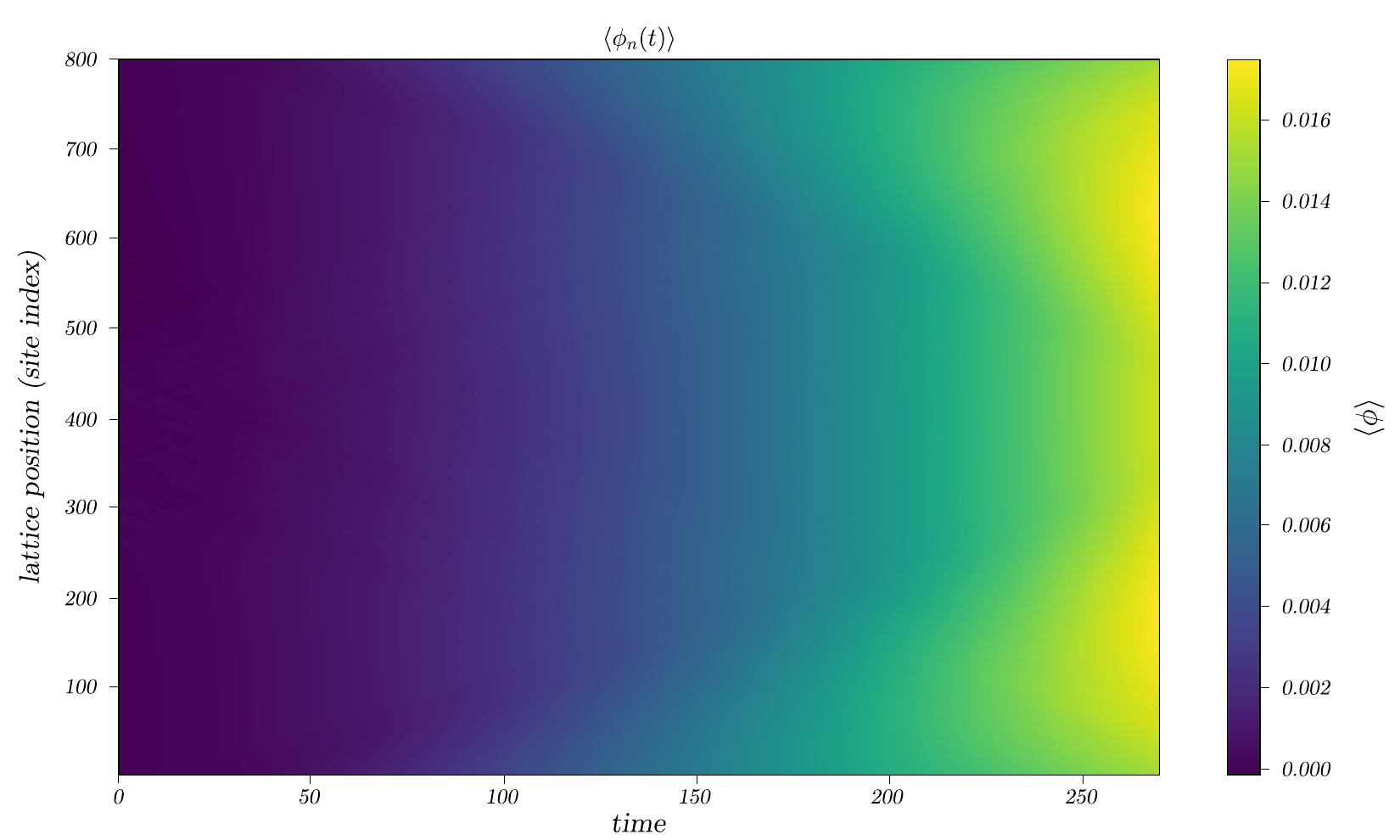}
        
        \vspace{0.2cm}
        (b) $\mu^2 = -0.25945$
    \end{minipage}
    
    \vspace{0.5cm}
    
    \begin{minipage}{0.48\linewidth}
        \centering
        \includegraphics[width=\linewidth]{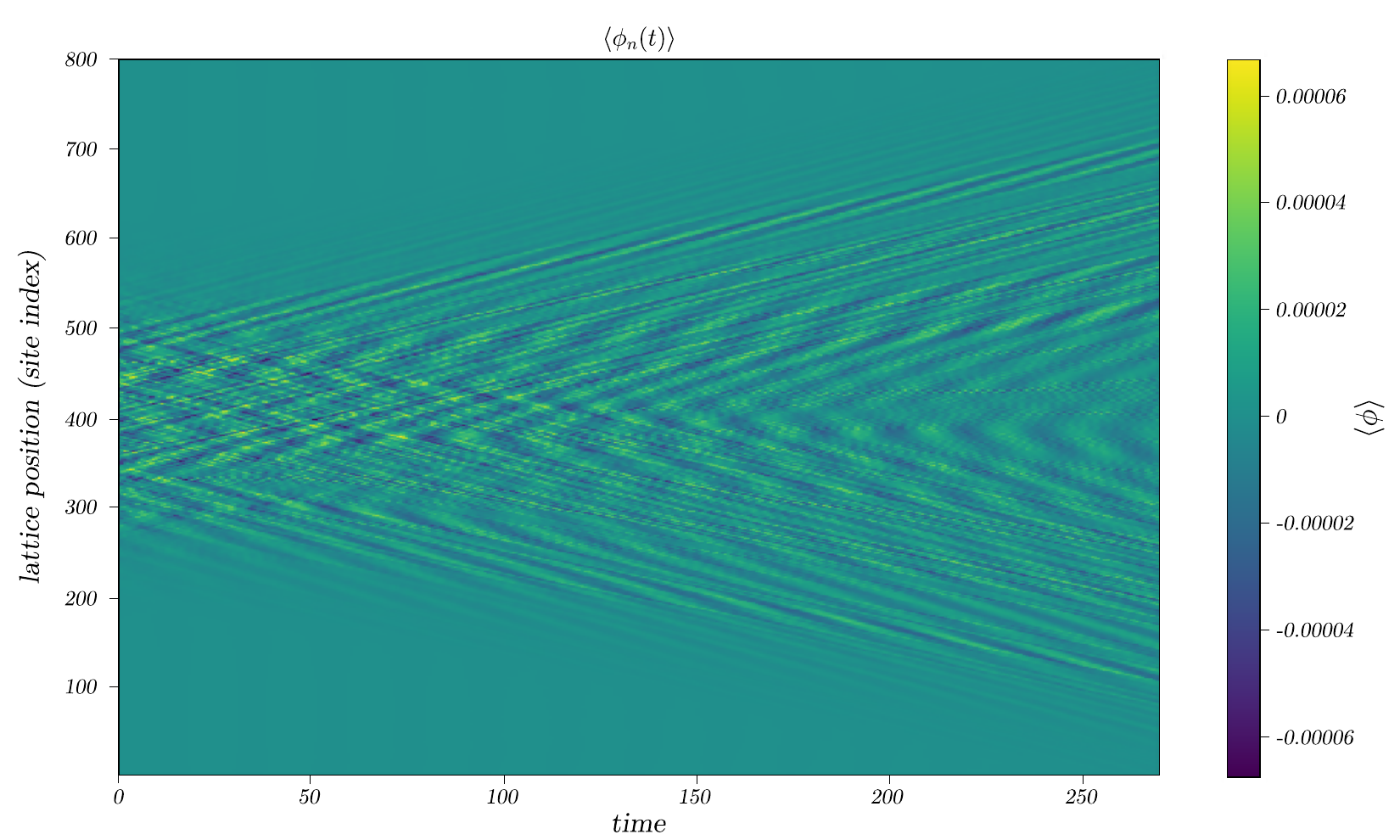}
        
        \vspace{0.2cm}
        (c) $\mu^2 = -0.1$
    \end{minipage}\hfill 
    \begin{minipage}{0.48\linewidth}
        \centering
        \includegraphics[width=\linewidth]{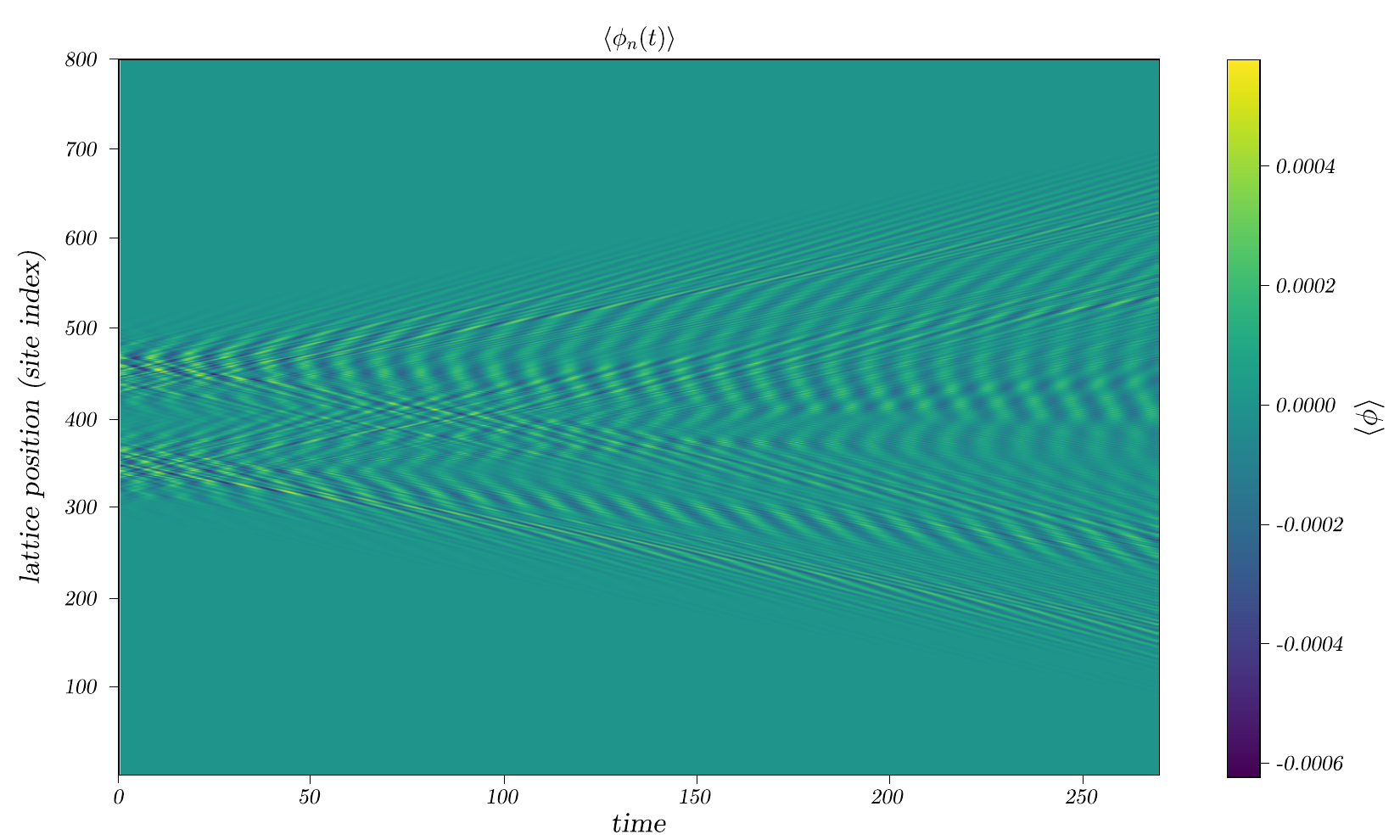}
        
        \vspace{0.2cm}
        (d) $\mu^2 = -0.5$
    \end{minipage}
    
    \caption{Real-time two-particle scattering in the $\phi^4$ theory at $\lambda = 0.8$, shown as space--time plots of $\langle \phi_n(t) \rangle$ in the sandwich geometry. Panels (a)--(d) correspond to the same values of $\mu^2$ as in Fig.~\ref{fig::spectra}. The gapped phases, (a), (c), (d), display the characteristic ``X'' pattern of wave packets approaching, colliding, and separating as outgoing quasiparticles, with strong inelasticity in the symmetric case (a) and almost perfectly elastic scattering in the broken phases (c), (d). At the critical point (b) the ``X'' pattern is absent and a slow drift of the profile indicates a breakdown of the sandwich protocol due to the diverging correlation length.}
    \label{fig::scattering}
\end{figure*}

To probe the nonperturbative dynamics we simulate real-time two-particle collisions across these regimes, as shown in Fig.~\ref{fig::scattering}. In the gapped phases, panels [\ref{fig::scattering}(a)], [\ref{fig::scattering}(c)], and [\ref{fig::scattering}(d)] display the characteristic diagonal ``X'' pattern in $\langle \phi_n(t) \rangle$, indicative of two localized wave packets approaching, interacting, and separating again as outgoing quasiparticles. The overall structure confirms the stable propagation of massive excitations in both the symmetric and spontaneously broken phases.

The detailed scattering outcome, however, depends strongly on the phase. In the symmetric phase [Fig.~\ref{fig::scattering}(a), $\mu^2 = +0.2$], the collision is highly inelastic: the extracted elastic probability and Wigner time delay at our reference energy $E_0$ are
\begin{align}
  P_{11\rightarrow 11}(E_0) &\simeq 0.712, 
  \\
  \Delta t_{\mathrm{sym}}(E_0) &\simeq -158.
  \label{eq:sym-prob-timedelay}
\end{align}
signaling a substantial cross section for particle production and other inelastic channels, as expected for a nonintegrable theory with a fundamental boson. By contrast, in the weakly broken phase [Fig.~\ref{fig::scattering}(c), $\mu^2 = -0.1$] the
scattering is almost perfectly elastic,
\begin{align}
  P_{11\rightarrow 11}(E_0) &\simeq 1, 
  \\
  \Delta t_{\mathrm{br1}}(E_0) &\simeq -108.2158,
  \label{eq:br-prob-timedelay}
\end{align}
with the incoming wave packets essentially re-emerging as outgoing packets with the same particle content and minimal radiation. The deeply broken point [Fig.~\ref{fig::scattering}(d), $\mu^2 = -0.5$] exhibits similar qualitatively elastic behavior, $P_{11\rightarrow11} \simeq 1$ and $\Delta t_{\mathrm{br2}}\simeq -177.781$, with even more clearly separated outgoing wave fronts due to the larger gap.

Near the critical coupling [Fig.~\ref{fig::scattering}(b)], the sandwich evolution ceases to produce a well-defined scattering event. Instead of the characteristic ``X''-shaped collision pattern, $\langle \phi_n(t) \rangle$ develops a long-wavelength drift across the entire window, and no clean separation into incoming and outgoing wave packets is observed. This is not a numerical artifact or an instability of TDVP/uMPS, but a direct consequence of criticality: the sandwich construction assumes a finite correlation length $\ell$ that separates a central scattering region from asymptotic bulk domains. At the critical point the exact ground state is gapless ($m=0$) and scale invariant, with a diverging correlation length ($\ell \to \infty$), so any finite-window implementation of this protocol inevitably violates that assumption and fails to reach asymptotic scattering states. In this sense, the breakdown of the sandwich geometry at $\mu^2 \approx \mu_c^2$ reflects the underlying physics of an infinitely correlated critical state rather than a flaw of the method, and it provides a dynamical signature of the quantum critical point that is complementary to the static FES diagnostics discussed above. Closer and closer to $\mu_c^2$ the required window sizes and evolution times grow with $\ell$, so in practice the protocol becomes rapidly more expensive and ultimately impractical as the critical point is approached.

\section{\label{sec::summaryDiscussion}Summary and Discussion}

We have extended TDVP-based uniform matrix product state techniques, previously applied to Ising field theory~\cite{Jha2025-IFT-MPS}, to the interacting $\phi^4$ quantum field theory in $(1+1)$ dimensions. A finite-entanglement scaling analysis at $\lambda = 0.8$ yields a controlled estimate of the critical mass-squared, $\mu_c^2 \in ]-0.2595,-0.2594[$, consistent with Ising universality and with the expected emergence of a gapless, conformal regime. This provides a quantitative map of the symmetric, near-critical, weakly broken, and deeply broken phases, anchoring the subsequent real-time simulations to a data-driven bracket of the critical point rather than to ad hoc parameter choices.

Using these uMPS ground states as asymptotic vacua, we simulated two-particle collisions in a sandwich geometry across the phase diagram and extracted both the elastic scattering probability $P_{11\to 11}(E)$ and the Wigner time delay $\Delta t(E)$. In the symmetric phase the collisions are strongly inelastic, with $P_{11\to 11} \simeq 0.712$, reflecting the sizable phase space for particle production in a nonintegrable theory with a fundamental boson. In the spontaneously broken phase the same protocol yields almost perfectly elastic collisions, $P_{11\to 11} \approx 1$, indicating that the excitations behave as stable, weakly radiating quasiparticles. As the system is tuned deeper into the broken regime, the increasing gap sharpens this behavior and further suppresses inelastic channels.

Near the critical coupling $\mu^2 = \mu_c^2$ the sandwich evolution fails to produce a clean scattering event: instead of a well-defined ``X'' pattern in $\langle \phi_n(t)\rangle$, we observe a long-wavelength drift across the entire window and no clear separation into incoming and outgoing packets. This breakdown is a direct manifestation of criticality: the scattering construction relies on a finite correlation length to isolate a central interaction region from asymptotic bulk domains, an assumption that is invalidated when the mass gap closes and $\xi \to \infty$. The failure of the sandwich protocol at $\mu^2_c$ thus provides a dynamical signature of the quantum critical point, complementary to the static FES diagnostics.

Taken together, these results demonstrate that uMPS--TDVP methods can resolve not only static spectra but also real-time scattering characteristics of an interacting relativistic quantum field theory across a phase transition. The combination of finite-entanglement scaling and real-time dynamics offers a general framework for locating critical points, characterizing quasiparticle content, and assessing the domain of validity of sandwich-geometry scattering constructions. Extensions to larger bond dimensions, refined wave-packet designs, and additional observables (such as detailed phase shifts, inelastic thresholds, and multi-particle production rates) should enable increasingly precise benchmarks of lattice $\phi^4$ theory and provide a template for applying tensor-network scattering techniques to more complex quantum field theories.

In future work it will be important to push the simulations to larger bond dimensions and local truncations, and to refine the construction of the incoming and outgoing states beyond the present diagonal-projection scheme, in order to obtain more precise phase shifts and time-delay profiles. The same framework can be extended to compare different scattering bases (e.g.\ total and relative momentum) and to incorporate improved analytical approximations for the quasiparticle content of the $\phi^4$ theory. These developments would turn the present work into the basis of a more comprehensive study of nonperturbative scattering and critical dynamics in interacting quantum field theories.

\section*{Acknowledgments}

We thank Raghav Jha for sharing his insightful comments on this work in private correspondence, and for his contributions to real-time scattering in Ising Field Theory, which served as a key inspiration for the present study.

W.C. gratefully acknowledges the support provided by Jens Eisert, including financial assistance from Freie Universität Berlin through a research fellowship funded by the Deutsche Forschungsgemeinschaft (DFG, German Research Foundation) within the Research Unit FOR~2724 ``Thermische Maschinen in der Quantenwelt'' (Fonds~0420611402). 

We also thank the Lebanese University for its continuous institutional support.

\providecommand{\noopsort}[1]{}\providecommand{\singleletter}[1]{#1}%

\end{document}